\documentclass[12pt]{article}

\usepackage{amsmath}
\usepackage{amssymb}
\usepackage{graphicx}

\def\lam{\lambda}
\def\Lam{\Lambda}

\def\dis{\displaystyle}
\def\nab{\nabla}

\def\a{\rm a}
\def\b{\rm b}

\begin{document}

\begin{center}
{\large\bf Cosmology in a Higher-Curvature Gravity}\\[8mm]
\end{center}
\hspace*{15mm}\begin{minipage}{13.5cm}
Y. Ezawa, H. Iwasaki, M. Ohmori, S. Ueda, N. Yamada and T. Yano$^{*}$\\[3mm]
Department of Physics, Ehime University, Matsuyama, 790-8577, Japan\\[1mm]
$^{*}$Department of Electrical Engineering, Ehime University, Matsuyama, 790-
8577, Japan\\[5mm]
{\footnotesize Email : ezawa@sci.ehime-u.ac.jp, hirofumi@phys.sci.ehime-u.ac.jp, seiji.ueda@nao.ac.jp, naohito@phys.sci.ehime-u.ac.jp and yanota@eng.ehime-u.ac.jp}
\\[8mm]
{\bf Abstract}

We investigated the cosmology in a higher-curvature gravity where the dimensionality
of spacetime gives rise to only quantitative difference, contrary to Einstein 
gravity.
We found exponential type solutions for flat isotropic and homogeneous vacuum 
universe for the case in which the higher-curvature term in the Lagrangian density 
is quadratic in the scalar curvature, $\xi R^2$.
The solutions are classified according to the sign of the cosmological constant, 
$\Lambda$, and the magnitude of $\Lambda\xi$.
For these solutions 3-dimensional space has a specific feature in that the solutions
are independent of the higher curvature term. 
For the universe filled with perfect fluid, numerical solutions are investigated 
for various values of the parameter $\xi$.
Evolutions of the universes in different dimensionality of spacetime are compared.\\

\end{minipage}

\section{Introduction}

Einstein gravity describes the observed universe fairly well.
It predicts, however, the presence of initial singularity in our universe
\cite{EH,Wald}.
This is usually interpreted to mean the limit of applicability of the Einstein 
gravity.
Candidates for the alternative theory come from both classical and quantum 
considerations.
In classical regime, there exists a possibility of avoiding the initial singularity 
in higher-curvature gravity(HCG) theories.
HCG seems to be natural when the gravity is very strong and the curvature is large,
since linearity in $R$ of the Lagrangian would be too restrictive.
Numerical investigations of the initial singularity in HCG have already been 
carried out \cite{Nariai-NT} and singularity free solutions have been found,
although the solutions are limited only to a very early stage of the universe.
HCG is also suggested from the consistency of quantum theory, 
e.g. quantum field theory in curved spacetimes or string perturbation theory.
In these cases tidal effects are important and equivalence principle would not work
effectively.

Thus HCG seems to work in both classical and quantum theory.
From the cosmological point of view, it may give a better description of the initial
stage of the universe than Einstein one and is expected to give a similar 
description of the later universe as the Einstein one since the curvature effects 
would then be small.
However, the whole evolution of the universe in HCG has not been investigated in 
detail so far and the description of the present state of the universe is not 
satisfactory\cite{BHV}.

There have been many investigations on cosmology in superstring motivated HCG 
theories.
They can be classified into two types.
In one of them, the results of the perturbation theory are used and the Gauss-Bonnet
combination is adopted.\cite{EM-KRT}
In another, a suitable hypothes is adopted to construct HCG Lagrangian density 
leading to produce non-singular cosmological solutions, instead of using 
the results of the perturbation theory.
\cite{MB,BEM,NO-O,Easson}
Both types of models yield interesting cosmological solutions, although they are not
satisfactory, since the solutions do not describe the present state of the universe.

Theoretically the Gauss-Bonnet combination has a nicely simple property.
However it is so simple that new generalized coordinates, coming from the time 
derivatives of the metric and inherent to HCG, do not appear in the canonical 
formalism at least for the Robertson-Walker spacetime.
Therefore it would be worthwhile to investigate other types of HCG.
One of the simple models is described by a function of the scalar curvature.
This type of models has long been investigated and includes many problems peculiar 
to HCG such as the definition of the new generalized coordinate, complicated 
constraint structure\cite{BL,EKKSY,Que} and conformal equivalence\cite{Conf}.
Thus it is interesting to see whether this type of models are applicable to describe
 the realistic evolution of the universe.

In this paper, we investigate the evolution of the universe in HCG, described by 
a function of the scalar curvature, using analytical and numerical methods.
In numerical analyses, we adopt the recent observational data as the initial data 
which leads to the accelerated expansion, instead of assuming  suitable initial data
 at $t=0$.
One of the main differences of HCG from Einstein one is that the dimension of 
the spacetime gives rise to only quantitative differences, e.g. while in Einstein 
gravity there is no dynamical degrees of freedom in (2+1)-dimensional spacetime, 
there remains such freedom in HCG \cite{EKKSY,Que}.
Therefore it is easy to compare spacetimes differing in dimensionality\cite{EOY}.
In other words it would be easy to transfer results obtained for lower dimensional
spacetime to higher dimensional one.
We investigated the universe for the case of vacuum universe and the universe 
filled with perfect fluid for various values of the coefficient $\xi$ of 
the $R^2$-term.
It is found that (3+1)-dimensional spacetime has somewhat peculiar property in 
the former case.

In section 2 we present the basic equations for general spacetimes as well as for 
homogeneous and isotropic universe in the case of $R^2$ gravity.
In section 3 analytical and numerical solutions are investigated.
We obtain exponentially evolving solutions for a particular model containing 
a term quadratic in scalar curvature.
The results of numerical investigations for the vacuum case and the universe filled 
with perfect fluid are presented for various values of $\xi$.
The solutions exhibit largely different behavior according to the sign of $\xi$. 
Section 4 is devoted to the summary and discussions.

\section{Formulation}

\subsection{Basic equations}

We consider a pure gravity type model in $(d+1)$-dimensional spacetime described by 
a Lagrangian density
$$
{\cal L}_{G}={1\over 16\pi \hat{G}}\sqrt{-\hat{g}}\,f(R)         \eqno(2.1)
$$
where $f$ is a differentiable function of the $(d+1)$-dimensional scalar curvature 
$R$ and a hat denotes a quantity defined in $(d+1)$-dimensional spacetime.
Field equations derived from (2.1) is known to be written as
$$
f'R_{\mu\nu}-{1\over2}\hat{g}_{\mu\nu}f
+\hat{g}_{\mu\nu}\Box f'-\nab_{\mu}\nab_{\nu}f'=0                  \eqno(2.2)
$$
where $R_{\mu\nu}$ is the Ricci tensor constructed from 
the $(d+1)$-dimensional metric $\hat{g}_{\mu\nu}$, $\nab_{\mu}$ is the covariant 
derivative with respect to $\hat{g}_{\mu\nu}$, $\Box\equiv \hat{g}^{\lam\rho}
\nab_{\lam}\nab_{\rho}$ and $f'\equiv df/dR$.

Equations (2.2) depend on the dimension of spacetime only implicitly.
The dependence becomes explicit in the canonical formalism.
The system described by the Lagrangian density (2.1) is highly constrained, so we 
follow the method of Buchbinder and Lyakhovich (BL) \cite{BL,EKKSY} for treating 
the higher-derivative Lagrangian. 
The Hamiltonian density ${\cal H}^{*}$ is obtained by the Legendre transformation of the modified Lagrangian density ${\cal L}_{G}^{*}$ which takes into account 
the constraints and the definitions of the generalized coordinates coming from 
the time derivatives of the metric by introducing the Lagrange multipliers.
${\cal H}^{*}$ has the following form:
$$
     {\cal H}^{*}_{G}=N{\cal H}_{G}+N^k{\cal H}_{k}+{\rm divergent\ terms}
                                                                   \eqno(2.3)
$$
where $N$ is the lapse function and $N^k$ is the shift vector.
${\cal H}_{G}=0$ and ${\cal H}_{k}=0$ are the Hamiltonian and momentum constraints,
respectively.
Choosing the coordinate system where the lapse function $N$ is equal to 1 and 
the shift vector $N^k$ vanishes, explicit form for ${\cal H}_{G}={\cal H}^{*}_{G}$ 
is given as follows\cite{EKKSY}:
$$
{\cal H}_{G}=2\Pi^{-1}\Bigl(p^{ij}p_{ij}-{1\over d}p^2\Bigr)+{2\over d}pQ
            +{1\over 2}\Pi R-\sqrt{h}f(R)
            -{d+1\over 2d}\Pi Q^2+\Delta \Pi-{1\over2}\Pi{\cal R}  \eqno(2.4)
$$
where $h_{ij}$ is the metric of the $d$-dimensional space, $p^{ij}$ the momentum 
canonically conjugate to $h_{ij}$, $h\equiv \det h_{ij}$, ${\cal R}$ the scalar c
urvature 
formed from $h_{ij}$.
$p\equiv h_{ij}p^{ij}$, $Q\equiv h^{ij}Q_{ij}$ where $Q_{ij}$ is the extrinsic 
curvature which is taken as new generalized coordinate replacing 
the time-derivative of $h_{ij}$.
$\Pi$ is the momentum canonically conjugate to $Q$.
The scalar curvature of the $(d+1)$-dimensional spacetime, $R$, should be  
expressed in terms of the canonical variables through a relation 
$\Pi=2\sqrt{h}f'(R)$.
It is seen that the dimension $d$ appears only in the coefficients of some of 
the terms in such a way that there is no special value of $d$.

\subsection{Homogeneous, isotropic and flat spacetimes}

In order to solve the field equations explicitly, we have to specify the model.
Here we specialize the function $f(R)$ to a quadratic function:
$$
f(R)=-2\Lambda+R+\xi R^2                                           \eqno(2.5)
$$
Then, in terms of canonical variables, $R$ is expressed as
$$
     R={1\over 2\xi}\Bigl(\kappa^2\Pi/\sqrt{h}-1\Bigr)             \eqno(2.6)
$$
where $\kappa=8\pi G$.
Furthermore we specialize the spacetime to the flat Robertson-Walker spacetime:
$$
ds^2=-dt^2+a(t)^2\tilde{h}_{ij}dx^idx^j                            \eqno(2.7)
$$
where $\tilde{h}_{ij}$ is the metric of flat $d$-dimensional static space.
Then ${\cal H}_{G}$ in (2.4) reduces to the following form:
$$
     {\cal H}_{G}={\kappa\over 8\xi}a^{-2}\Pi^2-{1\over4}\left(3Q^2+{1\over \xi}
                  \right)\Pi+{1\over 2}p_{a}aQ
                  +{1\over \kappa}\left(\Lambda+{1\over 8\xi}\right)a^2
                                                                   \eqno(2.8)
$$
where $p_{a}$ is the momentum canonically conjugate to the scale factor $a$.

For vacuum spacetime the canonical equations of motion are derived from (2.8):
$$
\left\{\begin{array}{l}
     \dis \dot{a}={1\over d}aQ,\ \ \ 
     \dot{p}_{a}={d\kappa\over 8\xi}a^{-(d+1)}\Pi^2-{1\over d}Qp_{a}
                -{d\over \kappa}\Bigl(\Lambda+{1\over 8\xi}\Bigr)a^{d-1},
\\[5mm]
     \dis \dot{Q}={\kappa\over 4\xi}a^{-d}\Pi
                 -{1\over2}\Bigl({d+1\over d}Q^2+{1\over \xi}\Bigr),\ \ \ 
     \dot{\Pi}={1\over d}\Bigl\{(d+1)Q\Pi-ap_{a}\Bigr\}
\end{array}\right.                                                 \eqno(2.9)
$$
From these equations, we have an equation for $a$
\footnote{Of course, (2.10) can be derived from the space components of (2.2).}:
$$
   \dis (d-1){\ddot{a}\over a}+{1\over2}(d-1)(d-2)\Bigl({\dot{a}\over a}\Bigr)^2
             -\Lambda
        + HCT=0                                                     \eqno(2.10)
$$
where $HCT$ represents the contribution of the higher curvature terms and is expressed as
$$
\begin{array}{ccl}
HCT&=&\dis d\xi\Bigl[\,4\Bigl({a^{(4)}\over a}\Bigr)
         +8(d-2)\Bigl({\dot{a}\over a}\Bigr)\Bigl({a^{(3)}\over a}\Bigr)
         +6(d-2)\Bigl({\ddot{a}\over a}\Bigr)^2
\\[5mm]
 &&  \dis \hspace{8mm}
       +2(3d^2-20d+21)\Bigl({\dot{a}\over a}\Bigr)^2\Bigl({\ddot{a}\over a}\Bigr)
        +{1\over2}(d-1)(d-4)(d-9)\Bigl({\dot{a}\over a}\Bigr)^4\,\Bigr].
\end{array}                                                        \eqno(2.11)
$$

\section{Solutions}

\subsection{Analytical solutions}

It is easily seen that (2.10) allows a particular solution of the form 
$a(t)=a_{0}e^{\lam t}$ 
with constant $a_{0}$.
Putting this form into (2.10), we have
$$
d(d-1)\lam^2-2\Lambda+\xi d^2(d+1)(d-3)\lam^4=0                    \eqno(3.1\a)
$$
so that
$$
\lam^2=\left[-d(d-1)\pm\sqrt{d^2(d-1)^2+8\Lambda\xi d^2(d+1)(d-3)}\right]
/2\xi d^2(d+1)(d-3)                                                \eqno(3.1\b)
$$
Thus the solutions are classified according to the value of
$$
\eta\equiv 8\Lambda\xi(d+1)(d-3)/(d-1)^2,                           \eqno(3.2)
$$
and the sign of $\Lambda$.
The solutions exist only for $\eta\geq -1$.
$$
\begin{array}{cllcl}
{\rm I}.&\dis  a(t)=a_{0}\exp{\Biggl[\pm\sqrt{{4\Lambda\over d(d-1)}}}\,t\,
\Biggr]&\ \ {\rm for}\ \ \eta=-1\ &{\rm and}\ &\Lambda>0
\\[7mm]
{\rm II}.&\dis a(t)=a_{0}\exp{\Biggl[\pm\sqrt{{4\Lambda\over d(d-1)\eta}
\Bigl(-1\pm\sqrt{1+\eta}\Bigr)}}\,t\,\Biggr]&
\ \ {\rm for}\ \ -1<\eta<0\ &{\rm and}\ &\Lambda>0
\\[7mm]
{\rm III}.&\dis a(t)=a_{0}\exp{\Biggl[\pm\sqrt{{4\Lambda\over d(d-1)\eta}
\Bigl(-1+\sqrt{1+\eta}\Bigr)}}\,t\,\Biggr]&
\ \ {\rm for}\ \ \eta>0\ &{\rm and}\ &\Lambda>0
\\[7mm]
&\dis a(t)=a_{0}\exp{\Biggl[\pm\sqrt{{4|\Lambda|\over d(d-1)\eta}
\Bigl(1+\sqrt{1+\eta}\Bigr)}}\,t\,\Biggr]&
\ \ {\rm for}\ \ \eta>0\ &{\rm and}\ &\Lambda<0
\end{array}                                                        \eqno(3.3)
$$
This classification can be made more directly (although not compact), e.g. 
$\eta=-1$ means $\Lambda\xi=-(d-1)^2/8(d+1)(d-3)$ or $\eta>0$ means 
$\Lambda\xi(d-3)>0$.
It should be noted that, for $d=3$, (3.1a) shows that there is no contribution from
the higher curvature term, so the solution is the same as in Einstein gravity.
This is also seen if the trace of (2.2) is taken and  (2.5) is used for 
$f(R)$.
Correspondingly, for $d=3$, and consequently $\eta=0$, only half of the type II and 
type III solutions have $\eta\rightarrow 0$ limit.
We note that, for $\Lam=0$, nontrivial solutions exist only for $d=2$ if $\xi>0$.
It would be expected that these exponential type vacuum solutions are relevant only 
to the early stage of the universe.
Thus, in order to examine whether the initial singularity of our universe can be 
avoided or not, other types of solutions are necessary.
They will be investigated numerically in the next subsection.

In later stage of the universe, we  assume the universe to be filled with perfect 
fluid, when the field equations are obtained by replacing the rhs of (2.2) by 
the energy-momentum tensor of the perfect fluid.
Equation for the scale factor $a$ is expressed as
$$
   \dis (d-1){\ddot{a}\over a}+{1\over2}(d-1)(d-2)\Bigl({\dot{a}\over a}\Bigr)^2
             -\Lambda
        +HCT=\kappa p.                                             \eqno(3.4)
$$

The field equations are supplemented by the equation of state
$$
p=\gamma \rho                                                     \eqno(3.5)
$$
where $p$ is the pressure, $\rho$ the total energy density and $\gamma$ a constant.
Energy-momentum conservation leads to a relation:
$$
\rho=\rho_{0}\Bigl({a\over a_{0}}\Bigr)^{d(\gamma+1)}.             \eqno(3.6)
$$
Analytical solutions are available only for radiation dominated era, $\gamma=1/d$, 
which reads
$$
a(t)\propto t^{2/(d+1)}.                                           \eqno(3.7)
$$
Also in this case, $d=3$ is exceptional in that the solution (3.7) is the same as 
in Einstein gravity.
Other solutions are investigated numerically in the next section.

\subsection{Numerical analysis}

In this section we present the results of the numerical analyses of the equation
(2.10), for the vacuum case, and (3.4) with (3.5), describing the universe filled 
with perfect fluid.
These equations were solved numerically by the 4th order Runge-Kutta method.

We start from the universe today and investigate the evolution of the universe 
both to the future and to the past.
The present values of the cosmological parameters are taken from the WMAP results
\cite{WMAP}.
The energy densities of matter $\rho_{M\,0}$ and dark energy (cosmological 
constant) $\rho_{DE\,0}$ are taken to satisfy
$$
({\rm i})\ \ \rho_{M\,0}\ :\ \rho_{DE\,0}=27:73,\ \ \ {\rm and}\ \ \ 
({\rm ii})\ \ \Omega_{M\,0}+\Omega_{DE\,0}=1\ \ \ ({\rm flat\ universe}).\eqno(3.8)
$$
The latter means, of course, the total energy density is critical.
The Hubble constant is taken to be
$$
H_{0}=72\;{\rm km\,s^{-1}Mpc^{-1}}.                                  \eqno(3.9)
$$
The value of $\ddot{a}_{0}$ is determined approximately from the Einstein equation 
and $a^{(3)}_{0}$ from the Hamiltonian constraint.
It is noted that these values of parameters lead to the accelerating universe as is 
shown in the following figures.

Figure 1 shows  the predicted evolutions of the universe filled with perfect fluid
for $d=3$ according to Einstein gravity and HCG for various values of the parameter 
$\xi$.
For positive $\xi$, the evolutions predicted by HCG are almost the same as that by 
Einstein gravity as expected.
On the other hand, for negative $\xi$, predictions of HCG are very different from 
that of Einstein gravity and depend heavily on the values of $\xi$.
These results remain unchanged for the value of $H_{0}$ other than the one in (3.9).
\footnote{The older values of these parameters\cite{Per,Rie,TSS,Efs,Free} lead to 
almost the same results.}.

Figure 2 shows the evolutions of the scalar curvature for the cases as in Figure 1.
The curvature becomes very large at early stage as expected and corresponds to 
the behavior of the scale factor.
The results shown in Figures 1 and 2 seem to rule out the cases of negative $\xi$ 
since the calculated age of the universe is so short.

Figure 3 shows the evolution of the scale factor for the dimensions of the space,
$d=2,3,4$ in Einstein gravity.
As to the initial conditions for $d=2$ and $d=4$, we tentatively adopted the same 
ones as for $d=3$.
Figure 4 shows the corresponding evolution of the scalar curvature.

Figures 5 and 6 show the same evolution as in Figures 3 and 4 in HCG for $\xi=10$.
The difference in the equation of motion is only quantitative as noted above, which 
is in accordance with our numerical results. 
We note that our numerical solutions satisfy the Hamiltonian constraint fairly well.

\begin{center}
\noindent
 \includegraphics[width=.8\hsize]{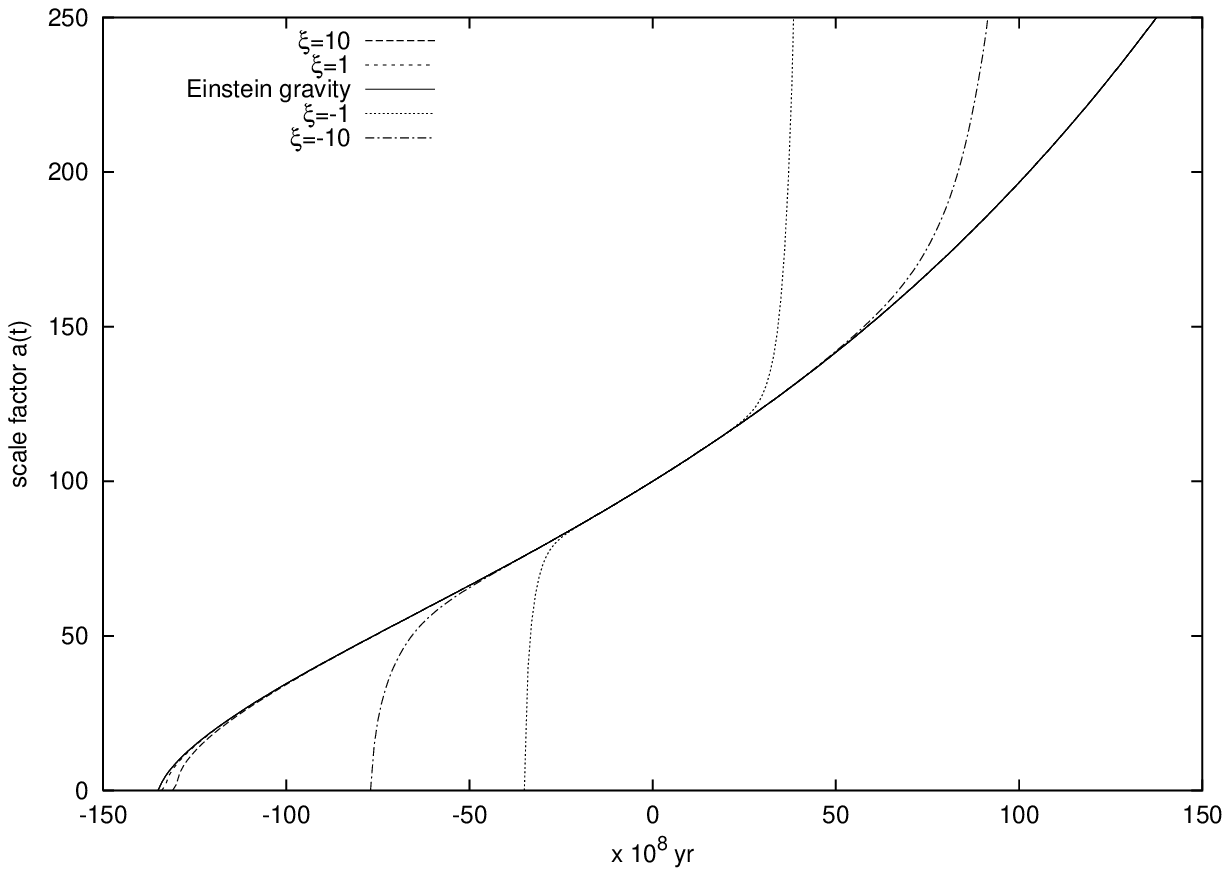}

\vspace{3mm}
\begin{minipage}{.7\hsize}
{\footnotesize Fig.1 Evolutions of the (3+1)-dimensional universe filled with 
perfect fluid in Einstein gravity and HCG for various values of $\xi$ }
\end{minipage}

\vspace{10mm}
\noindent
 \includegraphics[width=.8\hsize]{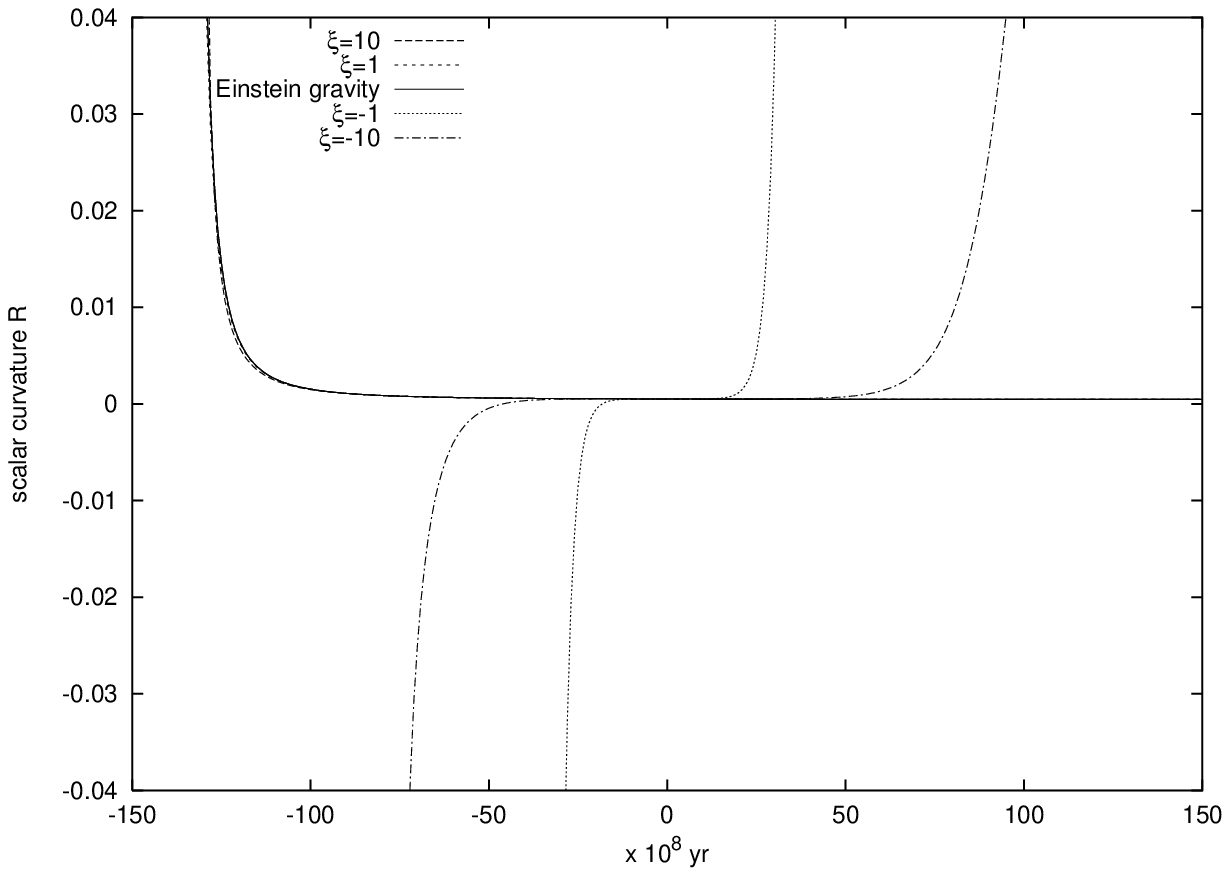}

\vspace{3mm}
\begin{minipage}{.7\hsize}
{\footnotesize Fig.2 Evolutions of scalar curvature $R$ in (3+1)-dimensional 
spacetime }
\end{minipage}

\end{center}

 \vspace{10mm}

\noindent
 \includegraphics[width=.5\hsize]{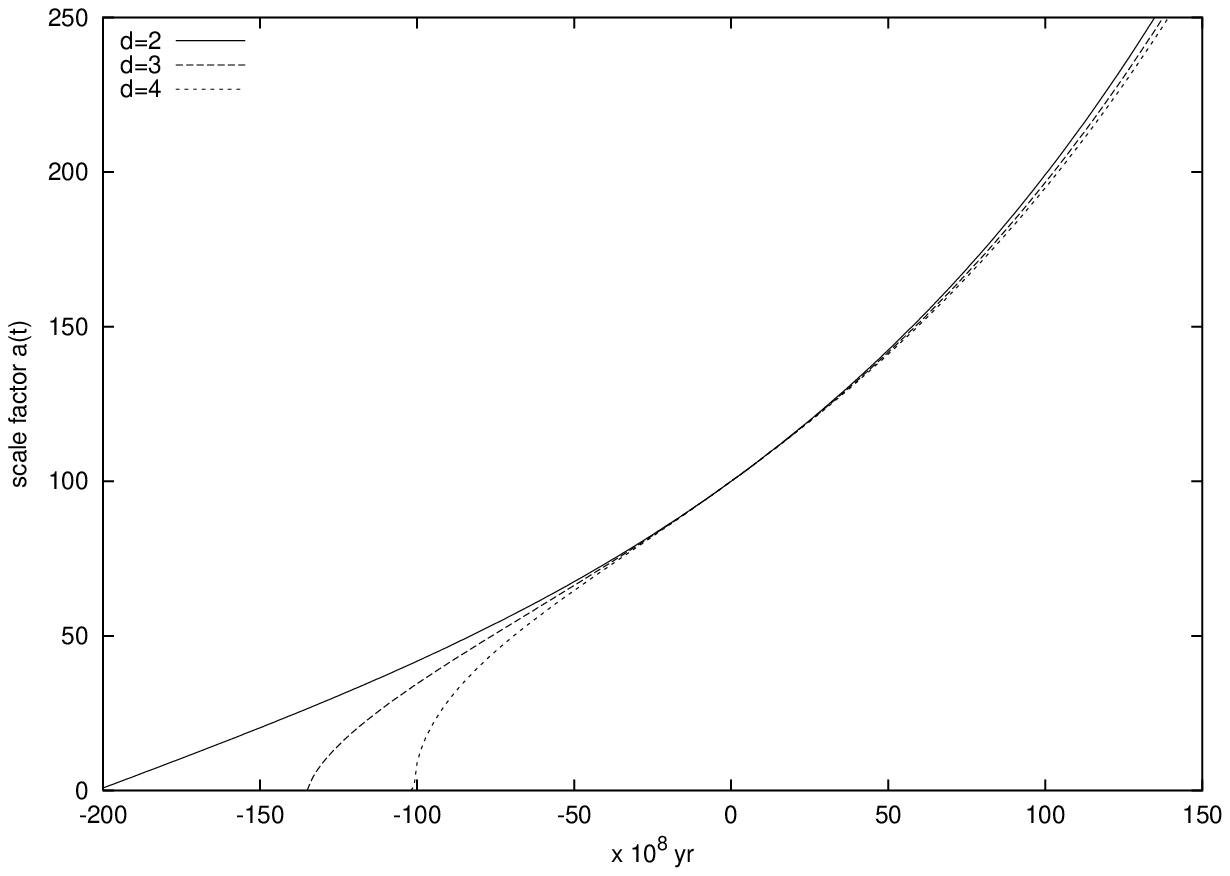}
 \includegraphics[width=.5\hsize]{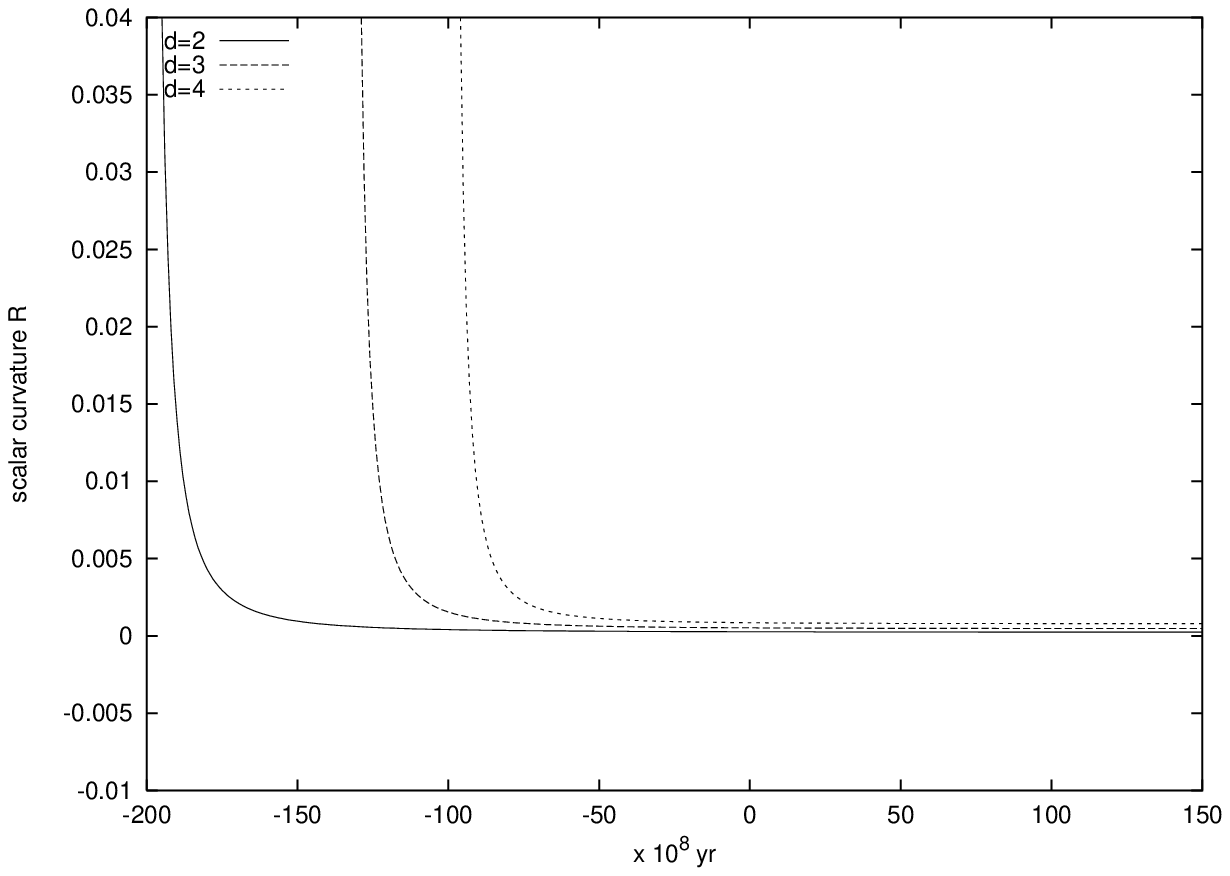}

\vspace{2mm}
\begin{minipage}{.45\hsize}
{\footnotesize Fig.3 Evolutions of the universe in Einstein gravity for dimensions 
$d=2,3$ and 4.}
\end{minipage}

\vspace{-12mm}
\begin{flushright}
\begin{minipage}{.45\hsize}
{\footnotesize Fig.4 Evolutions of scalar curvature $R$ in Einstein gravity for 
dimensions $d=2,3$ and 4.}
\end{minipage}
\end{flushright}

\vspace{10mm}

\noindent
 \includegraphics[width=.5\hsize]{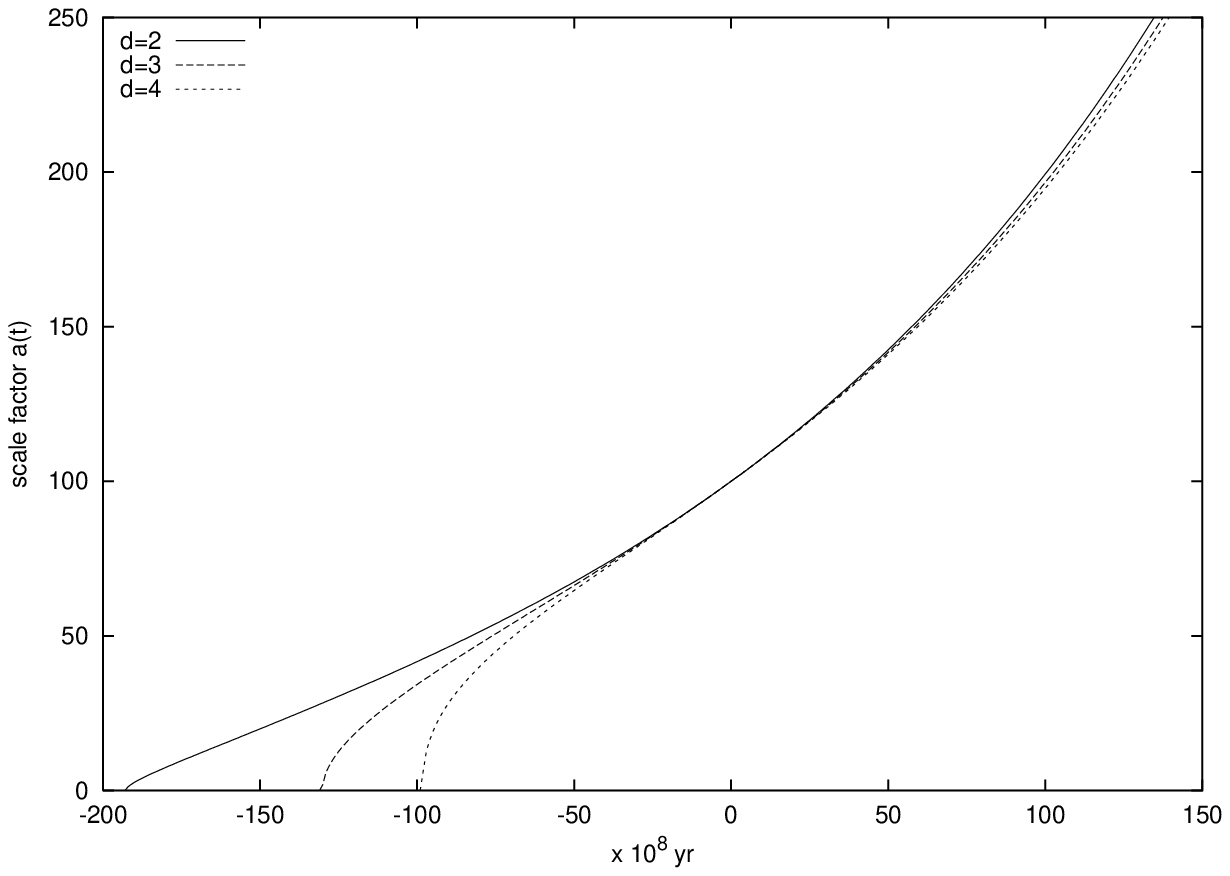}
 \includegraphics[width=.5\hsize]{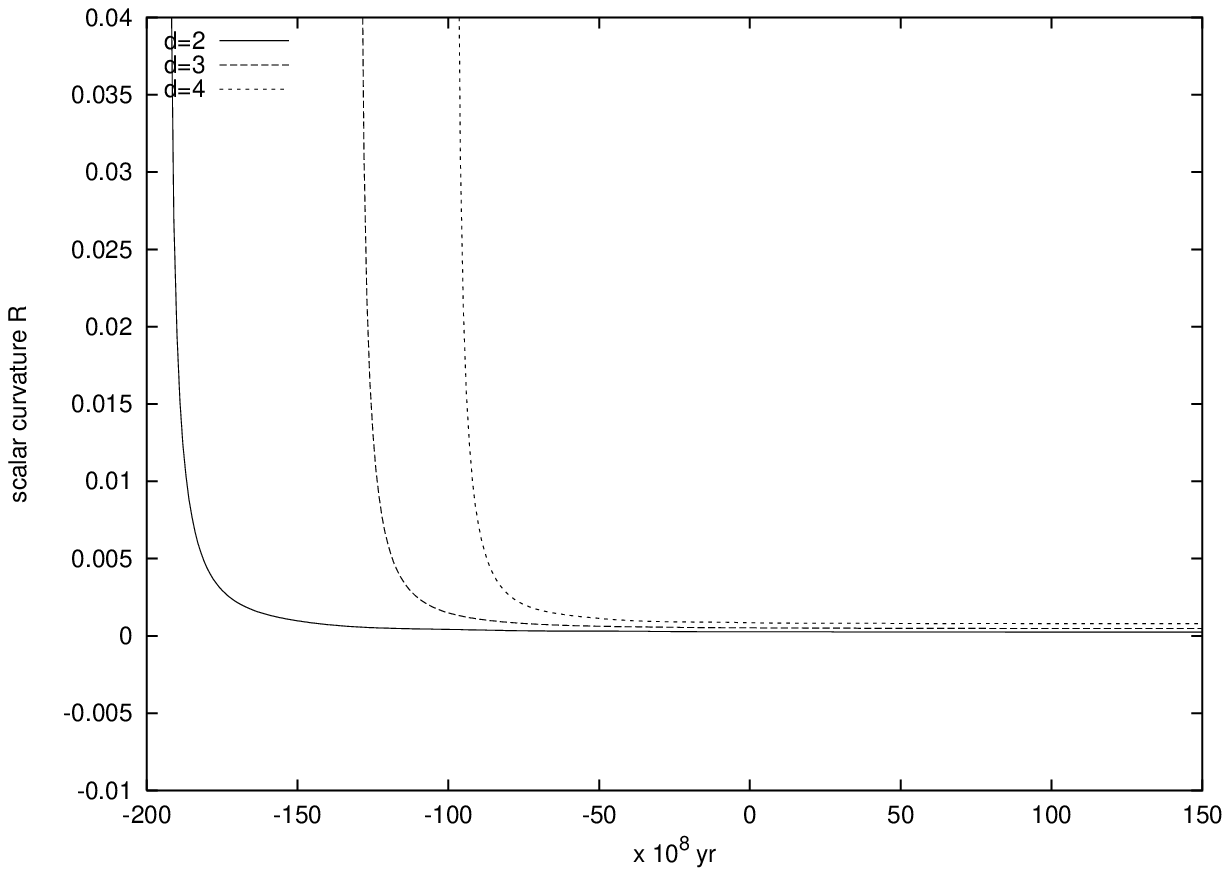}

\vspace{2mm}
\begin{minipage}{.45\hsize}
{\footnotesize Fig.5 Evolution of the universe in HCG with $\xi=10,\ $for dimensions $d=2,3$ and 4
 }
\end{minipage}

\vspace{-12mm}
\begin{flushright}
\begin{minipage}{.45\hsize}
{\footnotesize Fig.6 Evolutions of scalar curvature $R$ in HCG for $\xi=10$ for 
dimensions $d=2,3$ and 4}
\end{minipage}
\end{flushright}

\section{Summary and discussion}

We investigated the evolution of the flat, RW universe in a higher curvature gravity
(HCG) theory described by a Lagrangian depending on the scalar curvature.
Dimensionality of spacetime affects the evolution only quantitatively.
Explicit solutions are obtained when the higher curvature term is quadratic in 
the scalar curvature, $\xi R^2$.
Solutions exhibiting exponential expansion are obtained for the vacuum universe.
Exceptional case is that of 3-dimensional space, in which the solution is the same 
as in Einstein gravity, i.e. the higher curvature term gives rise to no effect.
These solutions may be interpreted to mean that other type of solution is necessary 
to avoid the initial singularity since the universe has not expanded exponentially 
throughout its evolution.

Numerical investigations were made by taking the recent observational data as 
initial conditions.
The solutions were compared for various values of the parameter $\xi$ and Einstein 
gravity.
For positive $\xi$, solutions in HCG deviate only slightly from that in Einstein 
gravity.
For negative $\xi$, solutions depend heavily on the value of $\xi$ and indicate too 
short an age of the universe.
However, numerical results do not show indications of avoiding the initial 
singularity in the range where our numerical calculations are valid.
It may be that a single set of equations is not sufficient to describe the whole 
history of the universe and different set of equations are required for the earliest
 and later stages of the universe analogous to the usual inflationary scenario.
In other words, it may be required that numerical solutions should be connected to
solutions, e.g. those obtained in the previous section at some early time.

Concerning the initial singularity, it is known that for spacetime with physically 
reasonable properties, occurrence of 
the singularity is inevitable if the Ricci tensor satisfies 
$R_{\mu\nu}\zeta^{\mu}\zeta^{\nu}\geq 0$ for any timelike vector 
$\zeta^{\mu}$.
In HCG with $\xi R^2$ term, we have
$$
\begin{array}{ccl}
R_{\mu\nu}\zeta^{\mu}\zeta^{\nu}&=&\dis
{1\over 1+2\xi R}\left[\kappa\left(T_{\mu\nu}\zeta^{\mu}\zeta^{\nu}
+{1\over d-1}T\right)-{2\over d-1}\Lambda\right]
\\[5mm]
&&\dis
-{\xi\over (d-1)(1+2\xi R)}\Bigl[R^2+2\Box R
         -2(d-1)\zeta^{\mu}\zeta^{\nu}\nab_{\mu}\nab_{\nu}R\Bigr].
\end{array}                                                        \eqno(4.1)
$$
The second term on the rhs is due to the higher curvature effect.
It is difficult to determine the sign of the rhs generally, even if the strong 
energy condition is satisfied.
It is seen that the results may change according the sign of $\xi$.
Restricting to the isotropic and flat universe filled with pressureless perfect 
fluid, we have, in terms of the canonical variables
$$
\begin{array}{ccl}
R_{\mu\nu}\zeta^{\mu}\zeta^{\nu}&=&
\dis {\sqrt{h}\over\kappa \Pi}\left[(u_{\mu}\zeta^{\mu})^2-{1\over d-1}\right]\rho
-{2\over d-1}\Lambda\\[5mm]
&&\dis -{1\over 2(d-1)\xi}\left[{\kappa^2\Pi\over\sqrt{h}}
+{\sqrt{h}\over\kappa^2\Pi}-2\right]
-{\Box\Pi\over (d-1)\Pi}+{\zeta^{\mu}\zeta^{\nu}\nab_{\mu}\nab_{\nu}\Pi\over\Pi}.
\end{array}
$$
Even in this case the situation is hardly improved so that numerical analysis seems 
to be necessary.
Our numerical results show that $R$ takes large positive value for $\xi>0$.
This indicates the possibility that the second term on the rhs of (4.1) dominates 
the first term and the numerical solutions could be connected to some non-singular 
solution.

We also compared the evolution of the universe for the dimensions of space $d=2,3\ 
{\rm and}\ 4$.
Intuitively the case of lower dimensional spacetime would seem to be simpler
\cite{EOY}.
However we found that the dimension of the spacetime is almost irrelevant in HCG.

Finally we comment on the conformal transformations in HCG\cite{Conf}.
In this work we identified the metric $\hat{g}_{\mu\nu}$ in (2.1) as physical and 
applied the initial conditions to them.
If the transformed metric is identified to be physical, the initial conditions are 
changed, although the equations of motion are only transformed.
The transformation is suggested by the form of ${\cal H}_{G}$ in (2.4) which is so 
complex.
The form can be brought to a simpler form.
Investigation of the transformed equations would be interesting since new types of 
solutions could be obtained.\\

\noindent
{\bf Acknowledgements.}
The authors thank the Yukawa Institute for Theoretical Physics
  at Kyoto University. Discussions during the YITP workshop
  YITP-S-02-01 on "The 25th Shikoku-seminar"
  were useful to complete this work.

\end{document}